\documentclass[a4paper]{article}

\usepackage[english]{babel}
\usepackage[utf8x]{inputenc}
\usepackage[T1]{fontenc}

\usepackage[a4paper,top=3cm,bottom=2cm,left=3cm,right=3cm,marginparwidth=1.75cm]{geometry}

\usepackage{braket}
\usepackage{amsmath}
\usepackage{amssymb}
\usepackage{physics}
\usepackage{graphicx}
\usepackage[colorinlistoftodos]{todonotes}
\usepackage[colorlinks=true, allcolors=blue]{hyperref}
\usepackage{slashed}
\def\inv{^{-1}}

\title{The Free Fermion Anomaly and Representations of the Pin Groups}
\author{Yakov Landau}

\begin{document}
\maketitle

\begin{abstract}
We consider a duality between a boson on a ring and a free fermion and show that they have an anomaly which corresponds to the states transforming under double covers of O(2). There are two (in general not isometric) double covers of O(2), known as Pin$_+(2)$ and Pin$_-(2)$. These can in general be distinguished at the group level by checking whether reflections square to $\pm1$. We show that in irreducible representations in complex Hilbert spaces the commutation of time reversal and charge conjugation gives another method for distinguishing Pin$_+(2)$ and Pin$_-(2)$. While we only demonstrate the duality for a single fermion, the anomaly is also present in any number of free fermions. For an even number of fermions we show that the two double covers of O(2n) are isomorphic. For an odd number of fermions we show that the distinction between irreducible representations of Pin$_+$ and Pin$_-$ can still be detected by the commutation of time reversal and charge conjugation namely Pin$_\pm(2n)$ will have $TCT = \pm C$  \end{abstract}

\section{Introduction}

Anomalies are a powerful tool primarily used in the context of quantum field theory to address questions about strongly coupled systems. Traditionally, continuous anomalies in even dimensions (associated to chiral fermions) are the most familiar ones (see~\cite{harvey2005tasi} for a review). The applications of discrete anomalies have been recently actively explored. The consequences of discrete anomalies are often as constraining as  those of the more familiar continuous anomalies. For some recent work see \cite{witten2016parity, cordova2018time, witten2016fermion, seiberg2016duality, hung2018linking, tachikawa2017gauging, numasawa2017mixed, guo2018time, cherman2017critical, tanizaki2018anomaly, cordova2018global, yamazaki2017relating, tanizaki2017circle, gomis2018phases, thorngren2017topological, kitano2017theta, di2017spontaneous, gaiotto2018time, kikuchi2017global, metlitski2017intrinsic, komargodski2018symmetry}. Discrete anomalies arise both in even and odd dimensions, and they clearly do not require the presence of chiral fermions (or fermions whatsoever). 
Here we study in detail a toy example of some quantum mechanical systems which have such discrete anomalies.
Many concepts familiar from the theory of continuous anomalies in even dimensions will arise in our quantum mechanical examples. For instance, we will encounter central extensions. The central toy example we consider is the theory of a free complex fermion and the theory of a boson on a ring with half-integer magnetic flux through the center of the ring. This is a toy example of bosonization duality. The ground states of the models agree and so do the discrete anomalies, as we check in detail.  Some aspects of the anomaly on the fermionic side was discussed in \cite{elitzur1986origins}, where it was argued that in the fermionic theory one cannot activate a chemical potential while maintaining charge conjugation symmetry. \\

This however is not sufficient to distinguish the two possible anomalies, which correspond to Pin$_-$ and Pin$_+$ structures. We resolve this ambiguity here. Furthermore, we carry out the same analysis in the bosonic theory and find exactly the same anomaly, corroborating the infrared duality.\footnote{Infrared duality in Quantum Mechanics means for us the same ground state structure and the same symmetries and anomalies.} (The bosonic version of the model we will discuss here appeared in connection with Yang-Mills theory in \cite{gaiotto2017theta},  and was presented at the 2017 Simons Center summer workshop by Dr. Jaume Gomis.) \\

We also discuss the anomaly of the system of several complex fermions and distinguish the Pin$_+$ versus Pin$_-$ structures of the anomaly in this case. 
We do not propose a natural non-Abelian version of the boson-fermion duality above, this remains an open problem and we hope that our work will lead to progress in this direction.
In addition, to make better contact between the theory of a boson on a ring and Yang-Mills theory, it would be nice to do a careful appropriate reduction of Yang-Mills theory to the quantum mechanical model and compare in detail the original discrete anomaly (which involves time reversal symmetry and one-form symmetry) with the anomalies of the particle on a ring (which involve charge conjugation, rotation, and time reversal symmetry). \\

The outline of the paper is as follows, we introduce the duality between the boson on a ring and a free fermion, followed by a discussion of the free fermion anomaly in full generality. We then introduce the pin group and classify its possible irreducible representations. In particular we focus on its irreducible representations on complex Hilbert spaces and show how classification can be accomplished using time reversal. We implement this classification to match the anomaly of the free fermion and the boson on a ring. We conclude by showing that this method can be used to classify the free fermion anomaly in full generality.

\section{Boson on a Ring -  Free Fermion Duality}

\subsection{Boson on a Ring}

Consider a one dimensional system consisting of a boson on a ring, with a magnetic flux going through the center of the ring. The Lagrangian for such a system is given by

$$\mathcal{L} = \frac{1}{2}\dot{q}^2 + \frac{\Theta}{2\pi} \dot{q}$$

The canonical momentum is $p = \dot{q} + \frac{\Theta}{2\pi}$, which tells us that the Hamiltonian for this system is given by 

$$\mathcal{H} = \frac{1}{2} \left(i\partial_q + \frac{\Theta}{2\pi}  \right)^2$$

We can consider momentum eigenstates in the q basis $\ket{p}$ with $\braket{q}{p} \sim e^{ipq} $. Since the boson is on a ring $q \sim q+2\pi$ and therefore $p = n \in \mathbb{Z}$. Normalization then gives us $\braket{q}{n} =\frac{1}{\sqrt{2\pi}}e^{iqn}$. These states are also energy eigenstates with $\hat{\mathcal{H}}\ket{n}= \frac{1}{2}\left(n-\frac{\Theta}{2\pi} \right)^2\ket{n}$. If we look at the Lagrangian we will see that it has translational symmetry, which on a ring becomes an SO(2) symmetry since $q\mapsto q+\alpha$ does nothing to the Lagrangian. This can be realized by an operator

$$V_\alpha \ket{n} = e^{i\alpha n}\ket{n} $$

For generic values of $\frac{\Theta}{2\pi}$ this is all we get. If $\frac{\Theta}{2\pi}$ is either an integer or a half integer then we get a degeneracy in the states. If $\frac{\Theta}{2\pi}$ is an integer then all states but the ground state are degenerate, and if $\frac{\Theta}{2\pi}$ is a half integer then every state is degenerate. We will focus on the half integer case where $\frac{\Theta}{2\pi} = \frac{1}{2}$. In this case we have a discrete $\mathbb{Z}_2$ symmetry which swaps the degenerate states 

$$C\ket{n} = \ket{1-n}$$

If we try and combine these two symmetries we do  not get an O(2) symmetry since we get

$$CV_\alpha C\ket{n} = e^{i\alpha}V_{-\alpha} \ket{n}$$

We can get an O(2) symmetry if rather than $V_\alpha$ as we defined it we centerally extend to $U_\alpha = e^{-i\frac{\alpha}{2}}V_{\alpha}$ in which case we do get the expected O(2) relationship $CU_\alpha C =U_{-\alpha}$. While we previously had $V_\alpha \sim V_{\alpha +2\pi}$ we now have $U_\alpha \sim U_{\alpha +4\pi}$, and so rather than having a representation of O(2) we see that the states are in a centrally extended representation corresponding to a double cover of O(2). \\

If we restrict our attention to the ground state, we have two states swapped by a $\mathbb{Z}_2$ operator with an associated double cover of O(2) as its symmetry group. This seems to be dual to a free fermion.

\subsection{Free Fermion}

A free fermion can be described by the Lagrangian

$$\mathcal{L} = i\psi^\dagger \dot{\psi}$$

where $\{ \psi^\dagger ,  \psi \} = 1$ and $\{ \psi ,  \psi \}  = 0 = \{ \psi^\dagger ,  \psi^\dagger \}$. This system can be expressed in terms of operators $\Gamma_1,\Gamma_2$ which satisfy $\{\Gamma_i,\Gamma_j\} = 2\delta_{ij}$, with $\psi = \frac{1}{2}(\Gamma_1+i\Gamma_) $ and $\psi^\dagger = \frac{1}{2}(\Gamma_1-i\Gamma_) $. Our Lagrangian can now be expressed as

$$\mathcal{L} = i \Gamma _i \dot{\Gamma}^i$$

This system clearly has an SO(2) symmetry as we can rotate the $\Gamma$s into each other and a $\mathbb{Z}_2$ symmetry from $\Gamma_i \mapsto -\Gamma_i$ which in this case do combine to give us an O(2) symmetry. However, given some rotation $R(\omega)$ which we try and implement via an operator $U(\omega)$ as 

$$\Gamma_i \mapsto {R_i}^j(\omega)\Gamma_j = U(\omega)\Gamma_iU^\dagger(\omega)$$

we see that $U(\omega)$ and $-U(\omega)$ implement the same transformation. So we again have the states transforming in a double cover of O(2). \\

To see if it is in fact dual to the boson on a ring we need to check that the extension of O(2) to its double cover is the same in both cases. There are in fact 2 double covers of O(2) called Pin$_+$(2) and Pin$_-$(2). These are not isomorphic so we need to see if the symmetry groups of the bosonic and fermionic systems are extended to the same double cover. \\

In general this duality only exists between the single free fermion and the boson on a ring. However the anomaly exists for all free fermion systems, and we'll try and understand the form of this anomaly in full generality.

\subsection{General Free Fermion Anomaly}

We consider the quantum mechanical system of $n$ free fermions described by the Lagrangian 

$$\mathcal{L} = i \psi^\dagger _\mu \dot{\psi}^\mu$$

where $\mu$ runs from $1$ to $n$ and the $\psi$s satisfy $\{ \psi^\dagger_\mu ,  \psi_\nu \} = \delta_{\mu \nu}$ and $\{ \psi_\mu ,  \psi_\nu \}  = 0 = \{ \psi^\dagger_\mu ,  \psi^\dagger_\nu \}$. We can instead take 2n $\Gamma_\mu$s with $\mu$ ranging from $1$ to $2n$, satisfying $\{ \Gamma_\mu, \Gamma_\nu \} = 2\delta_{\mu \nu}$. We can create raising and lowering operators  $\psi_\mu = \frac{1}{2} (\Gamma_{2\mu-1} - i \Gamma_{2\mu})$ and $\psi^\dagger_\mu = \frac{1}{2} (\Gamma_{2\mu-1} + i \Gamma_{2\mu})$. These will satisfy the above anticommutation relations.\\

With this relationship between $\Gamma$ and $\psi$ we can rewrite our Lagrangian as

$$\mathcal{L} = i \Gamma _\mu \dot{\Gamma}^\mu$$

for $\mu$ ranging from $1$ to $2n$. This Lagrangian has SO(2n) symmetry since we can rotate the $\Gamma$s into each other. We also have a $\mathbb{Z}_2$ symmetry since we can reflect any $\Gamma_\mu \mapsto -\Gamma_\mu$ and still leave the Lagrangian invariant. These combine to give us an O(2n) symmetry of the Lagrangian\\

We realize these symmetry transformations by some operators $U(\omega)$ which form a representation of O(2n) 

$$\Gamma_\mu \mapsto R(\omega)_\mu\ ^\nu \Gamma_\nu = U(\omega) \Gamma_\mu U^\dagger(\omega)$$

Again we see that for every $R(\omega)$ in the fundamental representation  $U(\omega) $ and $-U(\omega)$ implement the same transformation so this representation is at the very least a double cover of O(2n). The two double covers of the orthogonal group Pin$_\pm$(n) are in general not isomorphic.  We will study them and classify their irreducible representations on complex Hilbert spaces which will in turn classify the possible forms of this anomaly.

\section{Clifford Algebras and the Pin Group}

\subsection{Clifford Algebras}

We begin with a vector space $V$ over some field $K$ (in general not of characteristic 2 but we'll restrict to $K = \mathbb{R}$) and a quadratic form $Q:V \to K$. In general Q may be degenerate but we will restrict ourselves to cases where Q is non-degenerate. Associated to such a non degenerate quadratic form we can obtain a symmetric bilinear form $B:V \times V \to K$ by 

$$B(x,y) = \frac{1}{2} \left[ Q(x+y) - Q(x) - Q(y) \right] $$. 

The Clifford algebra associated with such a space is denoted as $C \ell (V,Q)$ and is an associative algebra which contains $V$ and can be constructed the following way.\\

Consider the tensor algebra over V, $T^\bullet V = \Sigma_k T^k V$ i.e. it is the collection of objects of the form $k_0 + k_1 v^{11} + k_2 (v^{21} \otimes v^{22}) + k_3 (v^{31} \otimes v^{32}\otimes v^{33}) + \cdots$ for $k_i \in K$ and $v^{ij} \in V$. Addition and multiplication in this algebra is given by tensor addition and multiplication. We then impose an equivalence relation on this space by asserting 

\begin{equation}
v\otimes v = Q(v)
\end{equation}  

The tensor algebra with this equivalence relation  is called the Clifford algebra. \\

\textbf{Note:} we made a choice of convention here by choosing $v\otimes v = Q(v)$, there are those that choose $v\otimes v = -Q(v)$. This choice is important for bookkeeping purposes, although it will obviously not effect the physics. \\

Addition and multiplication in this algebra is defined as in the tensor algebra, but with taking into account the equivalence relation we imposed (Equation 1). The Clifford product of $v,w \in  C\ell(V,Q)$ is  denoted $vw$. The equivalence $vv = Q(v)$ can also be expressed in another way. Consider $v,w \in V \subset C\ell(V,Q)$ then 

$$Q(v) + Q(w) + 2B(v,w) = Q(v+w) = (v+w)(v+w) = Q(v) + Q(w) + vw + wv$$ 

The first equality follows from the definitions of $B$ and $Q$, the second equality is an implementation of Equation 1 and the last equality follows from the definition of the Clifford product. This allows us to see that 

\begin{equation}
\{v,w\} = vw+wv = 2B(v,w)
\end{equation}

Given $v \in V \subset C\ell(V,Q)$ if $Q(v) \neq 0$ then we can define its inverse $v^{-1}$ as $v/Q(v)$ since   

$$v v^{-1} = v\frac{v}{Q(v)} = \frac{Q(v)}{Q(v)} = 1$$

The first equality follows from the definition of $ v^{-1}$ and the second by the definition of the Clifford product as defined in Equation 1.\\
\\
\subsection{The Pin Group}

We define the group Pin(V,Q) as the subgroup of $C\ell(V,Q)$ generated by multiplying elements of the form $u_i \in V \subset C\ell (V,Q)$ with $|Q(u_i)| = 1$, this is to say

$$\textrm{Pin}(V,Q) = \{u = u_1 u_2 \cdots u_k \in C\ell(V,Q) \: |  u_i \in V \subset C\ell(V,Q) \: \textrm{and} \: |Q(u_i)| = 1 \: \}$$

To see how Pin is related to the orthogonal group we start by a assigning to each $u \in V \cap \mathrm{Pin}(V,Q)$ a linear operator on V via the adjoint map $Ad:V \cap \textrm{Pin}(V,Q) \to \mathrm{End}(V)$ by $w \mapsto Ad_w$. We define $Ad_w :V \to V$ by $v \mapsto (w^{-1}) v w$. So for any $w \in V \subset C \ell (V,Q)$ we get a linear map from $V$ to $V$ which (using Equation 2) we can rewrite this as 

$$Ad_w(v) = (w^{-1}) v w = \frac{w}{Q(w)}(-wv + 2B(v,w)) = -v + \frac{2B(v,w)}{Q(w)} w$$

We can recognize this as $-R_w(v)$ where $-R_w$ is the reflection about the direction perpendicular to $w$ in the plane spanned by $v$ and $w$. This tells us that in fact that $Ad_w$ is an orthogonal transformation with respect to the bilinear from B (or the quadratic form Q), so $Ad:V \cap \textrm{Pin}(V,Q) \to O(V,Q)$. \\

This is slightly more transparent if we let $v,w$ be orthonormal basis vectors $e_\mu , e_\nu$ and then we would get $-R_{e_\nu}(e_\mu) = -e_\mu +2\delta_{\mu \nu}e_\nu$. We define the twisted adjoint by $\widetilde{Ad_w} :V \to V$ by $v \mapsto (\tilde{w}) v w$ where $\tilde{w} = -w^{-1}$. This gives us $\widetilde{Ad_w} (v) = R_w(v)$ and so we get actual reflections.\\

We can extend the domain of $Ad$ and $\widetilde{Ad}$ to the entire Pin group via composition. If we now consider two vectors $u,w$ then we define $\widetilde{Ad_{uw}}$ via their successive action on $v$.  $\widetilde{Ad_{uw}}(v) = \widetilde{Ad_u} \circ \widetilde{Ad_w} (v) = R_u \circ R_w(v) = (\tilde{u})( \tilde{w}) v w u $. The Cartan Dieudonne theorem \cite{lawson2016spin} tells us that the any element of the orthogonal group can be written as the composition of at most n such reflections. Therefore for every element in the orthogonal group can be realized as $\widetilde{Ad_{uw}}$ for some $u \in $Pin$(V,Q)$\\

So we now know that Pin(V,Q) covers O(V,Q), it is easy to see that this is a double cover since $Ad_w(v) = (\widetilde{w}) v w = (\widetilde{-w}) v (-w) = Ad_{-w}(v)$. This shows us that $w$ and $-w$ in Pin(V,Q) map to the same element in O(V,q). It is also true that the kernel of $\widetilde{Ad}$ is exactly $\{\pm 1\}$ and so we get a double cover. We will not prove this fact here but any thorough treatment of the subject will demonstrate it \cite{figueroaspin}. \\

If Q is positive definite the we denote Pin$(V,Q) := \textnormal{Pin}_+(V)$, if it's negative definite then we denote Pin$(V,Q) := \textnormal{Pin}_- (V)$. These can be distinguished from each other in a very simple way. In both of them, if we choose some $u \in V \cap \textnormal{Pin}(V,Q)$ then its corresponding action on V is to send $u \mapsto -u$ and keeps its orthogonal complement unchanged. However if we look at $u^2$ then we can distinguish Pin$_+$ from Pin$_-$. In Pin$_+$ we have $u^2 = Q(u) = 1$ whereas in Pin$_-$ we have $u^2 = Q(u) = -1$.\\

\section{Complex Representations of Pin}

As we are aiming to do physics here, we want a representation of of the pin group on a complex Hilbert space. We will be able to classify all the irreducible representations of relevant pin groups, by first doing so for the Clifford algebras. In classifying the representations of Clifford algebras, first we must classify them as algebras. We will make use of two results, stated without proof (see \cite{figueroaspin}). If we consider the real Clifford algebras associated to quadratic forms with p positive eigenvalues and q negative ones we'll denote $C\ell (p,q)$. All we need to classify $C\ell (p,q)$ is $p-q$ mod 8 and $p+q :=d$. With this, the Clifford algebras follow the simple pattern in Table 1, where by $\mathbb{K}(m)$ we mean  $m$ x $m$ matrices with entries in $\mathbb{K}$.

\begin{table}[h]
\begin{center}
\begin{tabular}{|c|c|}
\hline 
$p-q $ mod $ 8$ & $C\ell (p,q)$ \\
\hline
\hline
$0,6$ & $\mathbb{H}(2^{\frac{d-2}{2}})$ \\
\hline
$1$ & $\mathbb{R}(2^{\frac{d-1}{2}})\oplus\mathbb{R}(2^{\frac{d-1}{2}})$ \\
\hline
$2,4$ & $\mathbb{R}(2^{\frac{d}{2}})$ \\
\hline
$3,7$ & $\mathbb{C}(2^{\frac{d-1}{2}})$ \\
\hline
$5$ & $\mathbb{H}(2^{\frac{d-3}{2}})\oplus\mathbb{H}(2^{\frac{d-3}{2}})$ \\
\hline

\end{tabular}
\caption{Clifford algebra classification}
\end{center}
\end{table}

We are only interested in Pin$_\pm$(2n) arising from  $C\ell (2n,0)$ or $C\ell (0,2n)$ and therefore we are only looking at $p-q = \pm2n $ and so we don't care about the cases where $p-q$ is odd  so the only cases which interest us are the first and third rows in Table 1.. Furthermore, for an even number of fermions then we are looking at either $C\ell(4m,0)$ or $C\ell(0,4m)$. These are are isomorphic since $4m = -4m \; (\mathrm{mod}\;8)$. The only case in which we can make any distinction between Pin$_\pm$ is when we have an odd number of fermions. From here on out we will only concern ourselves with these. By looking at the table we see that these cases are either real or quaternionic,  and so we will make use of the following theorem (stated without proof see \cite{brocker2013representations}).\\

For $\mathbb{K = R,H}$ there exists a unique (up to isomorphism) irreducible representation of the algebra $\mathbb{K} (m)$ and that is $\mathbb{K}^m$. It therefore follows that the unique irreducible representation of the Clifford algebras $C\ell (0,2n)$ and $C\ell (2n,0)$ are either real or quaternionic.\\

Given any irreducible representation of a Pin group we can immediately build it up to an irreducible representation of its Clifford algebra since the Pin group contains all the generators of the Clifford algebra. Since the Clifford algebras we are interested in have only one irreducible representation the corresponding Pin groups also have only one irreducible representation. It therefore follows that the unique irreducible representations of Pin$_\pm(2n)$ are the same as their associated Clifford algebra and are either real or quaternionic.\\

Now we know that we have an irreducible representation of Pin groups into a subset of $\mathbb{K}(n)$ where $\mathbb{K} \neq \mathbb{C}$ and we want an irreducible representation on a complex Hilbert space. To do this we must consider isomorphisms between real or quaternionic representations and complex ones. We will do this by simply considering real and quaternionic representations as complex representations with additional structure \cite{brocker2013representations}. We will need quaternionic and real structures which consist of an anti-linear map $J$. If $J^2 = 1$ then $J$ is a real structure whereas if $J^2 = -1$ then $J$ is a quaternionic structure.\\

Given an algebra A an n dimensional quaternionic representation of A is $(V,\rho)$  where V is an n dimensional quaternionic vector space and $\rho:A \to \mathrm{End}_{\mathbb{H}}(V)$. We have a natural isomorphism between $\mathbb{H}^n$ and $\mathbb{C}^{2n}$, and so we have an isomorphism between any n dimensional quaternionic vector space and any 2n dimensional complex vector space. Let's call this isomorphism $\phi$.  Given any 2n dimensional complex vector space $W$ we can easily make it a complex representation of A as well by using the isomorphism $\phi$. Namely we define $\tilde{\rho}:A \to \mathrm{End}_\mathbb{C}(W)$ by $\tilde{\rho} = \phi \rho \phi \inv$. \\

However, as is, the representations $(V,\rho)$ and $(W,\tilde{\rho})$ are not isomorphic since $(V,\rho)$ is $\mathbb{H}$-linear while $(W,\tilde{\rho})$ is only $\mathbb{C}$-linear. If however we endow $W$ with an equivariant quaternionic structure J, i.e. an anti linear map J such that $J^2 = -1$ and for $\forall a \in A$  $\tilde{\rho}(a) J  = J \tilde{\rho}(a)$ then $(W,\tilde{\rho})$ is $\mathbb{H}$-linear and so $(V,\rho)$ and $(W,\tilde{\rho})$ are isomorphic. Conversely, given any complex representation it is isomorphic to a quaternionic representation if we can find an equivariant quaternionic structure.\\

Given an algebra A an n dimensional real representation is $(V,\rho)$  where V is an n dimensional real vector space and $\rho:A \to \mathrm{End}_{\mathbb{R}}(V)$. Given a real representation we can construct a complex one by $V \mapsto W = \mathbb{C}\otimes_\mathbb{R}V$. We define $\tilde{\rho}:A \to \mathrm{End}_\mathbb{C}(W)$ by $\tilde{\rho}(z\otimes v) = z\otimes \rho(v)$.  As before, this representation is $\mathbb{C}$-linear and is therefore not isomorphic to our original representation. If however, we endow W with an equivariant real structure, i.e. an anti-linear map J such that $J^2 = 1$ and for $\forall a \in A$  $\tilde{\rho}(a) J  = J \tilde{\rho}(a)$ then $(W,\tilde{\rho})$ is $\mathbb{R}$-linear and so $(V,\rho)$ and $(W,\tilde{\rho})$ are isomorphic. Conversely, given any complex representation it is isomorphic to a real representation if we can find an equivariant real structure. \\

To make contact with Pin$_\pm$ consider the following. Suppose we had an irreducible representation of Pin$_+(n)$ on some complex Hilbert space. Since Pin$_+(n)$ can be generated by n elements $e_\mu$ we could generate $\rho(\mathrm{Pin}_+)$ by n elements $\rho(e_\mu)$. We would be able to see that it was a Pin$_+(n)$ representation by virtue of the fact that $\rho(e_\mu)^2 = +1$. As a complex representation however, it would be isomorphic to a Pin$_-(n)$ representation by simply $\rho(e_\mu) \mapsto i \rho(e_\mu)$ which would satisfy $\rho(e_\mu)^2 = -1$. However as we know the irreducible representations are real (or quaternionic) and so there must an equivariant real (or quaternionic) structure J such that $J\rho(e_\mu) = \rho(e_\mu)J$ and so we need more than just $\rho(e_\mu) \mapsto i \rho(e_\mu)$ to map an $\mathbb{H}$-linear representation to an $\mathbb{R}$-linear representation or vice versa.

\section{Anomaly matching}

We'll begin with the free fermion case. we're looking either at Pin$_+$(2) or Pin$_-$(2), and their irreducible representations are either $\mathbb{R}^2$ or $\mathbb{H}$. The underlying complex space space of $\mathbb{H}$ is $\mathbb{C}^2$ and $\mathbb{C} \otimes \mathbb{R}^2 =\mathbb{C}^2 $ as well. So we want a representation of the $\Gamma$s in $\mathbb{C}^2$. We know such a representation, namely the pauli matrices so we let $\Gamma_\mu = \sigma_\mu$ for $\mu = 1,2$. The SO(2) rotations are given by $U(\omega)=$ exp $( \omega i\sigma_3 )$ = $\exp ( \omega [\Gamma_1, \Gamma_2] )$. The single reflection can be accomplished by $C = i\sigma_1$ since $C\Gamma_1 C\inv = \Gamma_1$ and $C\Gamma_2 C\inv = -\Gamma_2$. Our reflection satisfies $C^2 = -1$ and therefore we have a $\mathbb{C}$-linear representation of Pin$_-(2)$. We know however that the irreducible representation of Pin$_-(2)$ is $\mathbb{H}$ and this $\mathbb{C}^2$ representation is only isomorphic to it if we have an equivariant quaternionic structure. In this case we have a natural anti-linear operator to consider, namely time reversal.\\

Time reversal is given by an anti-linear operator $T$ which must reverse all the spins which means that $T\sigma_i T\inv = -\sigma_i$ for $i = 1,2,3$. Since $\sigma_1, \sigma_3$ are real and $\sigma_2$ is imaginary we can conclude that $T = \xi \theta \sigma_2$, where $\theta$ is the complex conjugation operator, and $\xi \in U(1)$ is an arbitrary phase. We therefore get that $T^2 = \xi \theta \sigma_2 \xi \theta \sigma_2 = \xi  (-\sigma_2) \xi^*  \sigma_2 = -1$. $T$ is therefore a quaternionic structure. Since $T\Gamma_\mu T\inv = -\Gamma_\mu$ we can see that $T \exp ( \omega [\Gamma_1, \Gamma_2] ) T\inv = \exp ( \omega [\Gamma_1, \Gamma_2] )$. Furthermore $TCT\inv = Ti\sigma_1 T\inv = i\sigma_1 = C$ and so we have an equivariant  quaternionic structure given by time reversal, and so we have an irreducible representation of Pin$_-(2)$ representation.\\

We could've just as well have taken $C=\sigma_1$ in which case $C^2 = 1$ and gotten a Pin$_+$(2) representation. In this case however, while $T \exp ( \omega [\Gamma_1, \Gamma_2] ) T\inv = \exp ( \omega [\Gamma_1, \Gamma_2] )$ still holds, we now have $TCT\inv = T\sigma_1 T\inv =-\sigma_1 = -C$ and so T is no longer equivariant. Consider then $J=T\Gamma$ where $\Gamma = \Gamma_1\Gamma_2$. In this case $J\Gamma_\mu J\inv = \Gamma_\mu$ so $JCJ\inv = C$ and $J \exp ( \omega [\Gamma_1, \Gamma_2] ) J\inv = \exp ( \omega [\Gamma_1, \Gamma_2] )$, and so J is equivariant. Furthermore $J^2 = T\Gamma T \Gamma = T^2\Gamma^2 = (-1)^2 = 1$. Therefore J furnishes us with an equivariant real structure and we have an irreducible representation of Pin$_+(2)$.

Which tells that although in the groups Pin$\pm$(2) the distinguishing factor was whether $C^2 =\pm1$, in the irreducible representations of $C^2 =\pm1$ we get something more. We now see that in dealing with irreducible representations of Pin$\pm$(2) we can in effect distinguish them by introducing T satisfying $T^2 = -1$ and seeing how T and C commute. We get the following:

\begin{table}[h]
\begin{center}
\begin{tabular}{|c|c|c|}
\hline 
Pin$_+(2)$ & $C^2 = 1$ & $TCT = C$ \\
\hline
Pin$_-(2)$ & $C^2 = -1$ & $TCT = -C$ \\
\hline
\end{tabular}
\caption{Classification of single free fermion anomalies}
\end{center}
\end{table}

So we now go to the bosonic case in which we defined $C$ by $C\ket{n} = \ket{-n+1}$ which gives $C^2 = 1$. We define $T\ket{n} = e^{in\pi}\ket{-n+1}$ so that $T^2 = -1$. In this case we get a Pin$_+$(2) representation which we can see either by the fact that $C^2 = 1$ or by the fact that $TCT = C$. As with the fermion case we could also take $C\ket{n} = i\ket{-n+1}$ in which case we would've gotten a Pin$_-$(2) representation since $C^2 = -1$ and $TCT = -C$

\section{General Free Fermion Anomaly}

Now we can consider our anomaly in the general free fermion case, the Lagrangian is 

$$\mathcal{L} = i \Gamma _\mu \dot{\Gamma}^\mu$$

We can use the $\Gamma$s to create the generators of SO(2n) by $\Sigma^{\mu \nu } = \frac{1}{4} [\Gamma^\mu, \Gamma^\nu]$ so, if we have a rotation with some parameter $\omega$ then the rotation will be implemented by 

$$\Gamma_\mu \mapsto {R(\omega)_\mu}^\nu \Gamma_\nu = U(\omega) \Gamma_\mu U^\dagger(\omega) = \exp\left[ \frac{1}{2} \omega_{\rho \lambda} \Sigma^{\rho \lambda}\right] \Gamma_\mu \exp\left[- \frac{1}{2} \omega_{\rho \lambda} \Sigma^{\rho \lambda}\right] $$

where $R(\omega)$ is the fundamental of SO(2n). We want to extend this to O(2n) via a single reflection. We therefore look for an operator $C$ which implements $C\Gamma_\mu C^{-1} = \xi_\mu\,^\nu \Gamma_\nu$ where $\xi = $ diag(-,+,...,+). \\

We can do this by choosing $C = \Gamma_1 \Gamma $ where $\Gamma $ is the product of all $ \Gamma_i$s i.e. $ \Gamma =  \Gamma_1 \Gamma_2 \Gamma_3 ... \Gamma_{2n}$. $\Gamma$ has the properties $\Gamma^2 = (-1)^n = -1$ since in our case n is always odd, and $\Gamma \Gamma_\mu = - \Gamma_\mu \Gamma$. This gives us that $C^{-1} =\Gamma_1 \Gamma = C$, and when acting on some $\Gamma_\mu$ we get 

\[ C\Gamma_\mu C^{-1} =
  \begin{cases}
    -\Gamma_\mu       & \quad \text{if } \mu = 1\\
    \Gamma_\mu  & \quad \text{if } \mu \neq 1
  \end{cases}
\]

Since this has $C^{-1} =\Gamma \Gamma_1 = C$ we get that $C^2 = 1$ and so this is a $(\mathbb{C}$-linear) Pin$_+$(2n) representation. We construct the corresponding Pin$_-$(2n) representation by considering $C = i\Gamma_1\Gamma$. \\

To get irreducible representations of Pin$_\pm$ we'll need an equivariant real or quaternionic structure. Unlike the 1 fermion case time reversal is not equivariant (although it's still a symmetry) and therefore it cannot act as a real or quaternionic structure. However, we will be able to use time reversal to build a real or quaternionic structure.\\

 The irreducible representations of the Clifford algebra (and of the Pin groups) is either $\mathbb{R}(2^{d/2})$ or $\mathbb{H}(2^{\frac{d-2}{2}}) $. If we consider their associated complex spaces we see that  $\mathbb{R}(2^{d/2}) = \mathbb{R}(2^{2n/2}) = \mathbb{R}(2^{n})$ which, after complexification, becomes $\mathbb{C}(2^n)$ and similarly $\mathbb{H}(2^{\frac{d-2}{2}}) = \mathbb{H}(2^{\frac{2n-2}{2}}) = \mathbb{H}(2^{n-1})$ which when treated as a complex space of twice the number of dimensions also becomes  $\mathbb{C}(2^n)$. We know of a  representation of the $\Gamma$s in $\mathbb{C}(2^n)$ which we can express as

$$\Gamma_{2\mu-1} = \mathbb{I}\otimes\cdots \otimes\mathbb{I}  \otimes \sigma_x  \otimes \sigma_z\otimes \cdots \otimes \sigma_z \:\:\mathrm{and}\:\:
\Gamma_{2\mu} = \mathbb{I}\otimes\cdots \otimes\mathbb{I}  \otimes \sigma_y  \otimes \sigma_z\otimes \cdots \otimes \sigma_z $$

with the $\sigma_x$ or $\sigma_y$ is in the $\mu$th position. In this basis time reversal can be written as 

$$T = \theta \sigma_y \otimes \cdots \otimes \sigma_y$$

which gives us $T^2 = (-1)^n = -1$ since we're only dealing with odd n, and

\[ T\Gamma_\mu T\inv =
  \begin{cases}
    -\Gamma_\mu       & \quad \text{if } \mu = 1,2 \mod 4 \\
    \Gamma_\mu  & \quad \text{if } \mu  = 3,4 \mod 4
  \end{cases}
\]

Which means that $T[\Gamma_\mu, \Gamma_\nu]T\inv = \pm [\Gamma_\mu, \Gamma_\nu]$ depending on the exact values of $\mu$ and $\nu$, which is why T is not equivariant even though it's still a symmetry. Since T flips the sign of some $\Gamma$s and not others we will define operators $J_\pm$ which will be anti-linear and either flip the sign of all the $\Gamma$s or of none of them. We will do this by first defining 

$$M = \prod_{\substack{
   \mu = 1,2 \\
   \mod 4
  }} \Gamma_\mu ,  \:\:\: P = \prod_{\substack{
   \mu = 3,4 \\
   \mod 4
  }} \Gamma_\mu $$

From which we define

$$J_+ = TM, \:\:\: J_- = TP$$

which satisfy $J_\pm\Gamma_\mu J_\pm\inv = \pm \Gamma_\mu$ which gives us $J_\pm[\Gamma_\mu, \Gamma_\nu] J_\pm\inv =[\Gamma_\mu, \Gamma_\nu]$ and so it's equivariant with respect to rotations which leaves us only to check reflections. $J_\pm \Gamma\Gamma_1 J_\pm \inv = \pm \Gamma\Gamma_1$ and so by anti-linearity $J_\pm i \Gamma\Gamma_1 J_\pm \inv = \mp i \Gamma\Gamma_1$. \\

All this combines to tell us the following, if we have a Pin$_+$(2n) representation with $C = \Gamma\Gamma_1$ then only $J_+$ is equivariant, and conversely if we have a Pin$_-$(2n) representation with $C = i\Gamma\Gamma_1$ then we only have $J_-$. We can see whether $J_\pm$ furnishes a real or quaternionic structure by squaring it and we see that 

$$J_-^2 = T^2P^2 = (-1)^{\frac{n-1}{2}}\:\: \mathrm{ and } \:\: J_+^2 = T^2M^2 = (-1)^{\frac{n+1}{2}}$$

So if $n = 1 \mod 4$ then Pin$_+$(2n) is quaternionic and Pin$_-$(2n) is real, and conversely $n = 3 \mod 4$ then Pin$_+$(2n) is real and Pin$_-$(2n) is quaternionic. This exactly matches the results of Table 1, as it should. \\

This gives us the same feature we saw with Pin$_\pm$(2), i.e. while the original way we have been able to tell the difference between Pin$_+$ and Pin$_-$ is by looking at the group elements and seeing whether $C^2 = \pm1$, now given an irreducible representation of Pin$_+$ or Pin$_-$ we can distinguish them by seeing whether  $TCT = \pm C$. Since $T\Gamma\Gamma_1T = -T\Gamma\Gamma_1T\inv = \Gamma\Gamma_1$ and by anti linearity of $T$ we get that $Ti\Gamma\Gamma_1T = - i\Gamma\Gamma_1$. This means that in general $TCT = \pm C$ for Pin$_\pm$(2n), which gives us the two following ways of distinguishing irreducible representations of Pin$_\pm$

\begin{table}[h]
\begin{center}
\begin{tabular}{|c|c|c|}
\hline 
Pin$_+(2n)$ & $C^2 = 1$ & $TCT = C$ \\
\hline
Pin$_-(2n)$ & $C^2 = -1$& $TCT = -C$ \\
\hline
\end{tabular}
\end{center}
\end{table}

\section{Acknowledgements}
The author would like to thank Dr. Zohar Komargodski for suggesting this problem and for his invaluable input and guidance, Dr. Jaume Gomis for his input, and Dr. Martin Rocek for very helpful conversations and for being an all around amazing human being. The author would also like to thank the the Simons Center for Geometry and Physics where a portion of this work was done over the summer of 2017.

\bibliographystyle{alpha}
\bibliography{bibliography}

\end{document}